# ThreadPoolComposer – An Open-Source FPGA Toolchain for Software Developers

Jens Korinth, David de la Chevallerie, Andreas Koch
Embedded Systems and Applications Group (ESA)
Technische Universität Darmstadt
Darmstadt, Germany
Email: {jk, dc, ahk}@esa.cs.tu-darmstadt.de

***Abstract*—**
**This extended abstract presents ThreadPoolComposer, a high-level synthesis-based development framework and meta-toolchain that provides a uniform programming interface for FPGAs portable across multiple platforms.**

## I. Introduction

In recent years, major advances in *High-Level Synthesis (HLS)* have spawned a new generation of hardware compilers (such as LegUp [1] or Nymble [2] in the academic domain, or Xilinx Vivado HLS in industry) which can generate efficient, behaviorally equivalent hardware for computing kernels described in C/C++. Until recently, these tools were burdened not only with tackling the highly complex task of generating hardware from a C/C++ specification, but also with the equally daunting task of system synthesis, namely providing an entire hardware/software environment for the generated hardware kernels. This encompasses, e.g., defining and connecting to memories, managing host/FPGA communication and making the FPGA accessible using appropriate software interfaces.

ThreadPoolComposer aims to divide the task of *generating an FPGA hardware design* into the actual **HLS problem**, and the **problem of generating on-chip micro-architectures at the system level**. The main goals of ThreadPoolComposer are to provide *an easily customizable open-source tool suitable for researchers* investigating the latter problem, and *a common benchmark environment for researchers* working on HLS tools, while isolating software developers from the low-level mechanisms.

ThreadPoolComposer was developed in context of the EU FP7 research project REPARA [3], which aims for an automated front-to-back development flow for heterogeneous parallel computers encompassing one or more of multi-core, GPU, FPGA, and DSP-based processing elements. The flow can begin with legacy C++ code which is then incrementally refactored into modern C++11/14, from which in turn high-level code suitable for the different target processors can be deduced.

## II. ThreadPoolComposer

In the following, the ThreadPoolComposer toolchain and framework will be presented in a top-down approach, i.e., from the software interface down to the hardware bitstream generation for an FPGA device.

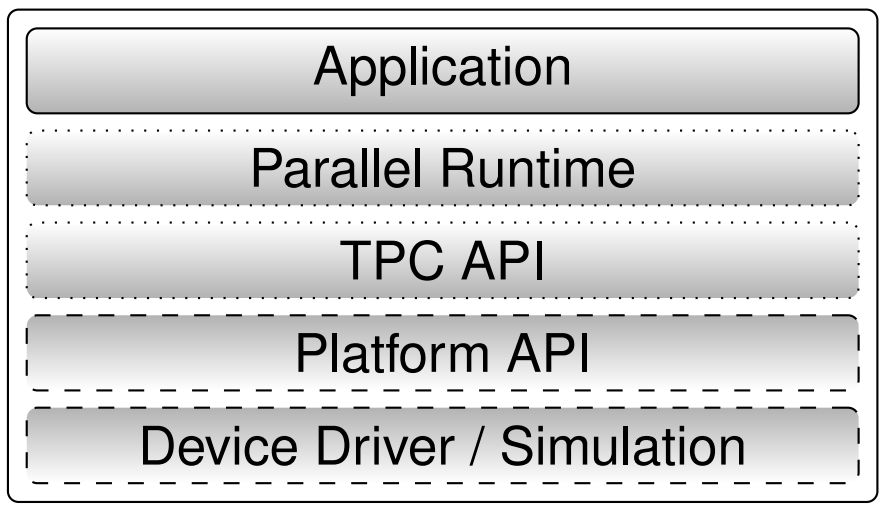

Fig. 1: ThreadPoolComposer Software Stack

### A. TPC API

The TPC API is the upper-most API layer (see Fig. 1); either the application directly uses TPC API, or it uses a parallel runtime framework (such as OpenCL, FastFlow [4]) which interfaces with TPC API. Its core tasks are 1) device enumeration and management 2) data transfer to and from the device 3) job preparation and launching. Listing 1 shows an example snippet of a job launch:

```
/* allocate 1 KB on device */
tpc_handle_t h = tpc_device_alloc(dev, 1024);
/* copy array 'data' to device */
tpc_device_copy_to(dev, data, h, 1024, TPC_BLOCKING_MODE);
/* prepare a new job for kernel id #10 (magic) */
tpc_job_id_t j_id = tpc_device_acquire_job_id(dev, 10);
/* set argument #0 to handle h */
tpc_device_job_set_arg(dev, j_id, 0, sizeof(h), &h);
/* launch job */
tpc_device_job_launch(dev, j_id, TPC_BLOCKING_MODE);
/* call blocks until completed, so get return value */
int r = 0;
tpc_device_job_get_return(dev, j_id, sizeof(r), &r);
printf("result␣of␣job:␣%d\n", r);
/* release job id */
tpc_device_release_job_id(dev, j_id);
/* release device memory */
tpc_device_free(dev, h);
```

Listing 1: TPC API Example

First, a small block of memory is allocated on the device via `tpc_device_alloc`, to which some data is copied via `tpc_device_copy_to`. Then, a job is requested and prepared by setting the first argument to the memory handle, i.e., the kernel shall work with the data that has just been transferred to the device. A job can be launched on the device either in *blocking* *or non-blocking mode*, i.e., the call returns after the job has finished, or immediately. Finally, the results are collected and the device memory is freed. This style is reminiscent of OpenCL, which was a deliberate choice to flatten the learning curve. Also note that we deliberately decided for manual data transfer management in order to give runtime schedulers optimization opportunities, e.g., by keeping intermediate results on the device between job executions. Such capabilities are currently being integrated into the FastFlow [4] run-time for heterogeneous parallel computers.

### B. Platform API

A wide range of FPGA-based processing platforms exists, ranging from reconfigurable systems-on-chip to larger PCI Express-based accelerators. Each device is usually aimed at a very specific audience and designed with certain applications in mind, which benefit hugely from the chosen architecture. This diversity cannot be easily unified without giving up a significant amount of the appeal of FPGA platforms. Therefore, the Platform API is inserted as a secondary software abstraction layer beneath TPC API; its purpose is to implement all *device-specific functionality*, currently: 1) device memory management 2) access to hardware registers, device memory 3) device-host communication and feedback. Both APIs can be implemented as *shared libraries*, giving the additional benefit of being *exchangeable at runtime*. From the software developer's perspective this allows moving between platforms **without recompilation of the application**. This facilitates *design space exploration* for any given application and increases re-usability.

### C. Compilation Toolchain

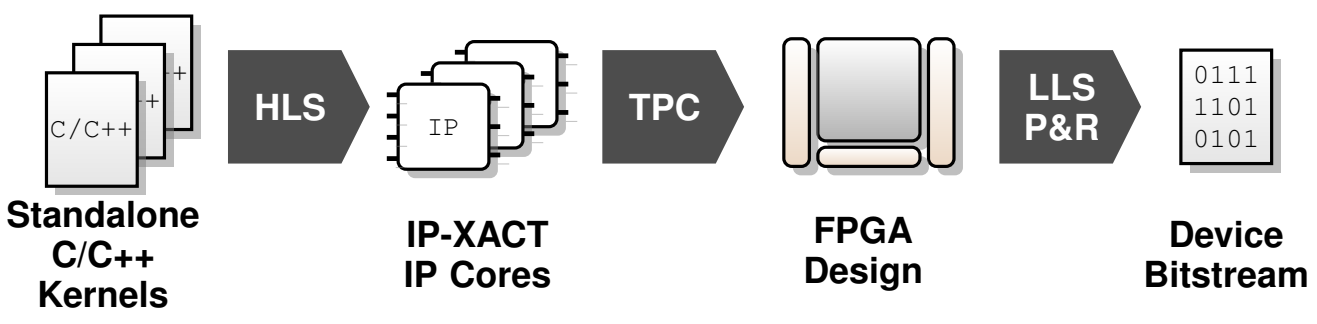

Fig. 2: TPC Compilation Flow

The overall compilation flow with ThreadPoolComposer is depicted in Fig. 2: Standalone C/C++ kernels have been extracted from the application and behaviorally equivalent IP cores are generated using *Xilinx Vivado HLS*. ThreadPoolComposer instantiates and arranges IP cores according to the given Composition, first creating a ThreadPool micro-architecture, which is wrapped in a Platform to yield a complete, synthesizable design. Finally, *Low-Level Synthesis (LLS)* is performed using the FPGA vendor toolchain (Xilinx Vivado).

The central idea behind ThreadPoolComposer is to *use HLS only as a C-to-Hardware compiler* at the level of individual accelerators, as opposed to being used as a *C-to-System compiler*, which would need to create an entire hardware system of accelerators, as well as internal and external interfaces, etc. While this is possible, it is rather awkward. Instead, the developer identifies and extracts computational kernels from the application (probably using tool support), then selects a Platform, i.e., a device or device family of FPGAs, and specifies a Composition to the ThreadPoolComposer toolchain: Such a Composition defines 1) the kernels to be used in the design 2) the desired number of parallel processing elements (PEs) for each kernel (i.e., the degree of parallelism for each kernel) and 3) an Architecture, i.e., a construction template for the organization of the PEs in the design. Fig. 3 illustrates the general organization of the design: The Architecture defines the template to instantiate an on-chip organization of PEs called the ThreadPool, which connects to the host and memory via an hardware infrastructure instantiated by a template provided by the chosen Platform. Note that the dependencies between the template types have been minimized to enable maximal re-use of existing Architectures on new Platforms. This also facilitates comparisons, e.g., of different Architectures on a given Platform. The toolchain is based on *Scala/SBT* and the structural templates are written in *Vivado IP Integrator Tcl*, which makes ThreadPoolComposer very easy to customize, modify, and extend.

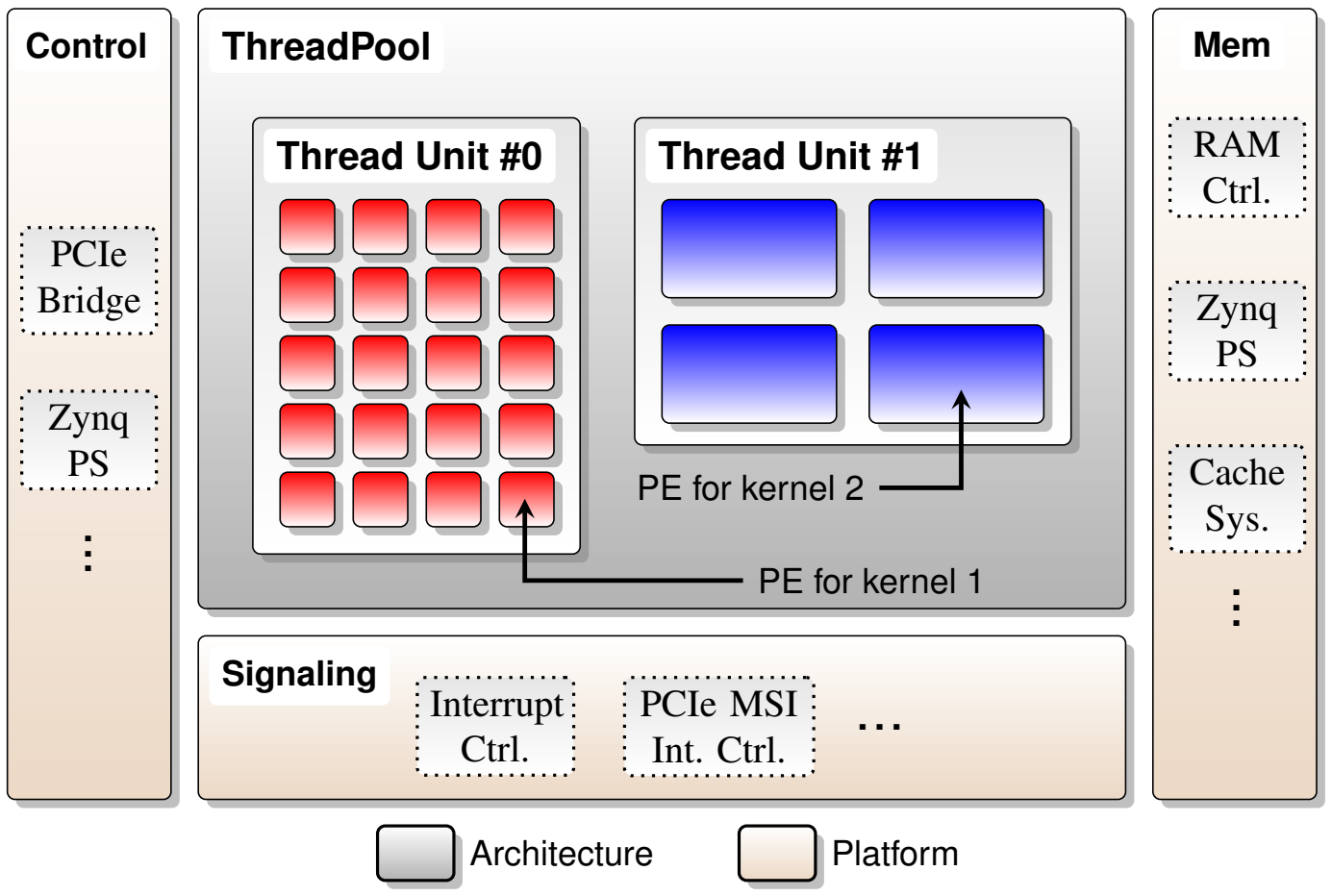

Fig. 3: FPGA Design Organization

## III. Platform Evaluation

ThreadPoolComposer is a work-in-progress and has not undergone thorough optimization yet. Currently, the system supports three Platforms using different classes of FPGA evaluation boards: The *zedboard* features a Zynq-7000 series XC7Z020-CLG484-1 FPGA with $F_{max}$ of 100 MHz and a dual-core ARM Cortex A9 at 666 MHz as host processor running Xilinx Linux 3.17.0. Xilinx' ZC706 is a larger version of the same system, using a XC7Z054-FFG900-2 FPGA with $F_{max}$ of 250 MHz and the same Cortex A9 running at 800 Mhz. Finally, the VC709 uses a 8x PCIe Gen3 interface based on **ffLink** [5] on a host with an eight-core Intel Xeon E5-1620v2 running at 3.7 GHz and Linux 3.19.5.

Fig. 4 shows the average data throughput in an otherwise idle system: Obviously, the VC709 benefits hugely from its PCIe interface, which transfers up to ≈4.2GiB/s at a chunk size of 512 KiB (and even more for much larger chunks, see [5]). The Zynq Platforms currently use kernel DMA buffers for the transfers, and their allocation leads to a significant slowdown. A *zero-copy approach* is currently under development to address this deficiency.

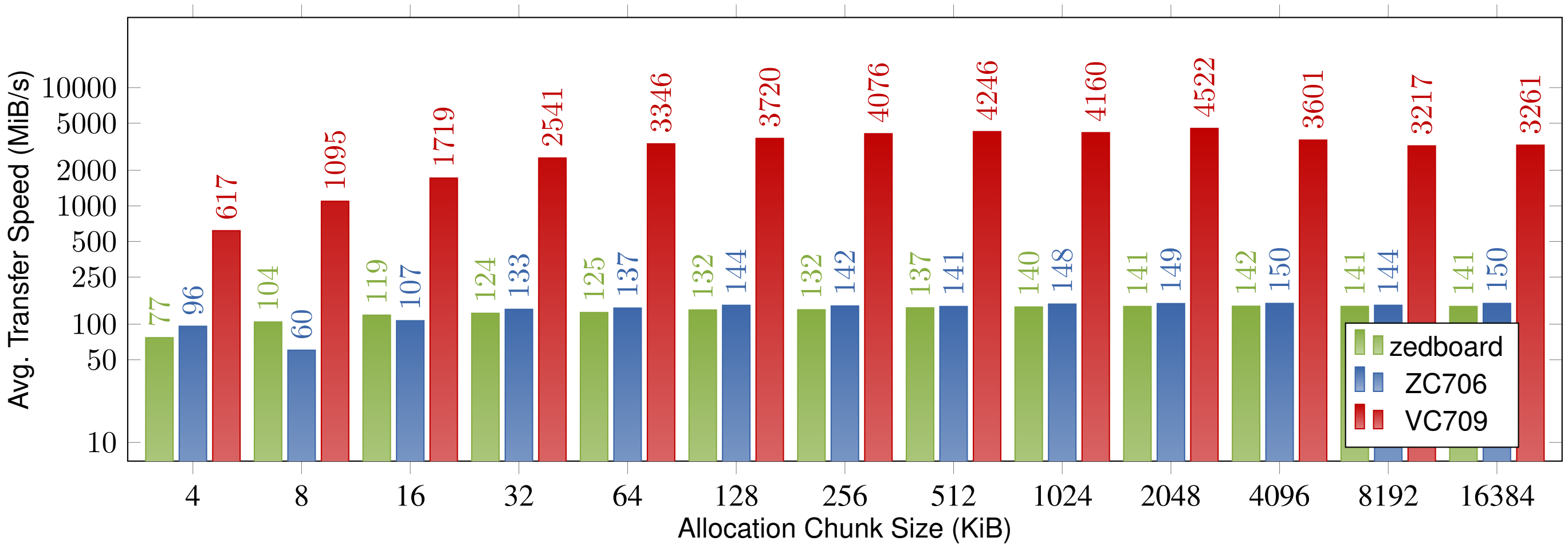

Fig. 4: Average bidirectional transfer rates between FPGA and host (i.e., user application memory) in MiB/s.

Fig. 5 depicts the interrupt latencies of the three platforms: To evaluate hardware/software round-trip time, we used a hardware counter to count the clock cycles between raising an interrupt (in hardware) and receiving the acknowledgement from the software (also in hardware). This measurement includes all intermediate software layers from OS level up to user application level. Latencies range from 3.2$\mu$s up to 22.5$\mu$s; shortest latencies can be achieved at the shortest kernel runtimes $\leq 10\mu s$ (calling thread is not put to sleep at all). Surprisingly, even though the VC709 has to transport interrupts via PCIe packets (and not dedicated wires), the latencies are significantly lower (almost by 2x). This is primarily due to the eight-core Xeon E5 running at 5x the speed of the ARMs on the zedboard.

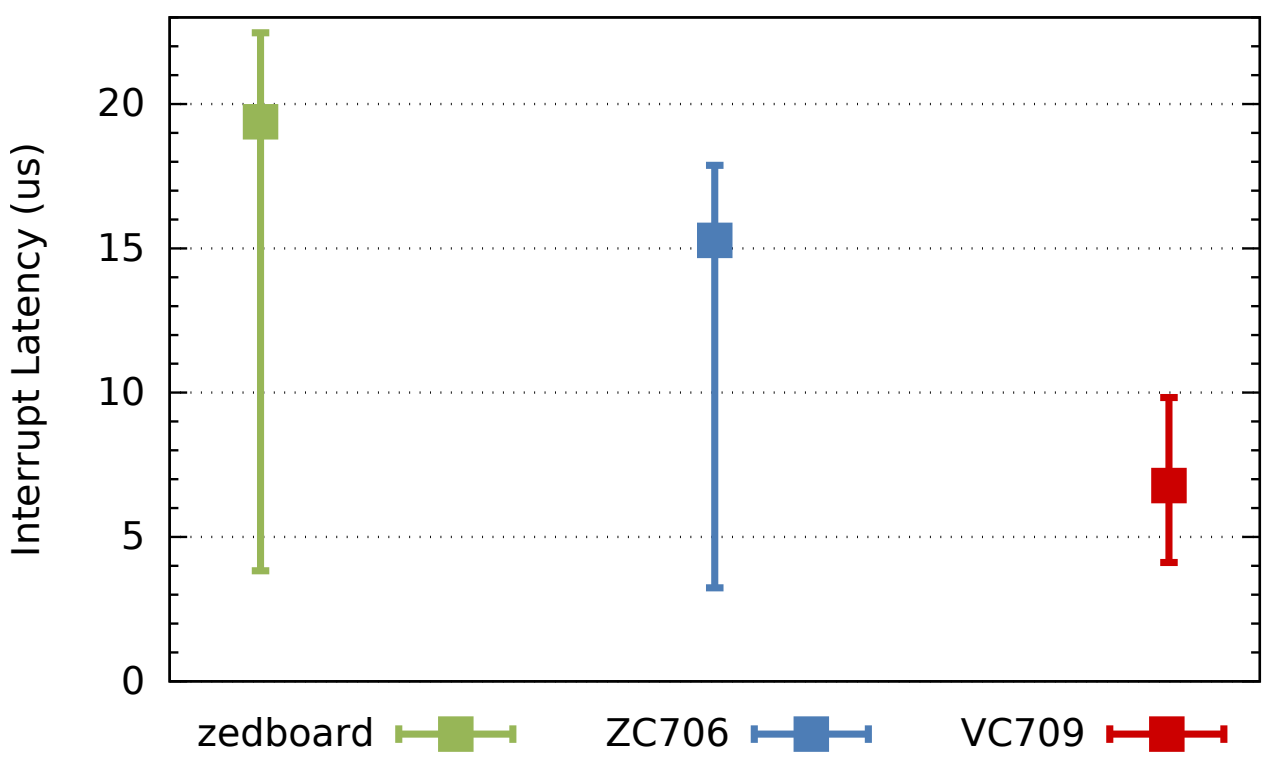

Fig. 5: Interrupt latencies: measured round-trip time (max/avg/min) between a hardware kernel signaling an interrupt and then receiving an acknowledgement from the host.

## IV. Conclusion & Future Work

ThreadPoolComposer is an open-source meta toolchain which facilitates the exploration on-chip microarchitectures for FPGA accelerators, comparison of HLS tools and separates HLS from system-on-chip architecture generation. It furthermore provides a unified API for software developers, which can be used with every combination of ThreadPoolComposer Platforms and Architectures, thus improving the separation of concerns, and provides a solid basis for future automated architecture exploration efforts. ThreadPoolComposer currently supports the Zynq and zedboard devices, as well as the Xilinx VC709 with PCIe Gen3 x8 support. In future work, we aim to further increase the performance of the hardware designs by developing custom IP cores and integrate them by on-the-fly hardware generation via the Chisel language [6].

*ThreadPoolComposer will be released in open-source form from the* Downloads *section of www.esa.cs.tu-darmstadt.de later this year.*